\documentstyle[12pt,axodraw,epsf]{article}

\if@twoside
 \else
 \fi
 
 \parskip \medskipamount
\newcommand{\be}{\begin{eqnarray}}
\newcommand{\ee}{\end{eqnarray}}

\begin{document}
\hbox{}
\nopagebreak
\vspace{-3cm}
\addtolength{\baselineskip}{.8mm}
\baselineskip=24pt
\begin{flushright}
{\sc  TPI--MINN--96--28/T} \\
{\sc    NUC--MINN--96--22/T} \\
{\sc    HEP--MINN--96--1524} \\
\end{flushright}

\begin{center}
{\Large \bf The BFKL Equation from the Wilson Renormalization Group}
\vspace{0.1in}

\vspace{0.5in}
{\large  Jamal Jalilian-Marian$^1$, Alex Kovner$^1$, Andrei Leonidov$^2$
and Heribert Weigert$^3$}\\
$^1${\it Physics Department, University of Minnesota\\
116 Church st. S.E., Minneapolis, MN 55455, USA}\\
$^2${\it Theoretical Physics Department, P.N.~Lebedev Physics Institute,\\
117924 Leninsky pr. 53  Moscow, Russia}\\
$^3${\it University of Cambridge, Cavendish Laoratory, HEP, Madingley Road,
Cambridge CB3 0HE UK}\\

\vspace{0.1in}

\vspace{1in}

{\sc  Abstract} 
\end{center}

\noindent
We discuss the Wilson renormalization group approach to the
effective action for low $x$ physics. It is shown that in the
linearized, weak field regime the RG equation reduces to the BFKL
equation for the evolution of the unintegrated gluon density.  We
discuss the relation of this approach with that of Lipatov.

\vfill

\newpage

\section {Introduction}

 Physics of dense gluon systems is one of the most rapidly growing
branches of quantum theory of strong interactions
(see  \cite{Le} for a recent review). The recent wave of interest
was sparked by experimental data on deep inelastic scattering \cite{HERA}
showing a considerable growth of gluon density in a proton at small Bjorken
 $x$. 
 
 Theoretical understanding of small $x$ physics is a fascinating
 problem in itself because it envolves a delicate blend of
 perturbative and nonperturbative QCD physics. On the one hand, due to
 large partonic densities, the QCD coupling constant in this region is
 expected to be small. On the other hand however, the large density
 must also lead to appearance of some collective effects (perhaps
 similar to the ones often disussed in the framework of the high
 temperature QCD) which should be taken into account
 nonperturbatively.
 
 The perturbative paradigm of the low $x$ gluon physics is the famous
 BFKL equation \cite{BFKL}.  It resums all contributions of the form
 $(\alpha_s \ln 1/x)^n$ in the naive perturbative series.  This is a
 linear equation, which describes the evolution with $x$ of the
 unintegrated gluon density $\varphi (y=\ln 1/x,{\bf k})$, related to
 the standard gluon distribution function $x g(x,Q^2)$ by
\begin{equation}
x g(x,Q^2) = \int_{0}^{Q^2} {d^2 k \over {\bf k}^2} \varphi (x,{\bf k})
\end{equation}
The equation is
\begin{equation}
{\partial \varphi (y,{\bf q}^2) \over \partial y} = {\alpha_s N_c \over \pi^2}
\int d^2 k\ K({\bf q}, {\bf k}) \varphi (y,{\bf k})
\label{BFKL}
\end{equation}
The action of the BFKL kernel on a function reads
$$
\int d^2 k\ K({\bf q},{\bf k}) \varphi (y,{\bf k})=
\int d^2 k\ \bigg[ K_{re}({\bf q},{\bf k})
+K_{virt}({\bf q},{\bf k})\bigg]\varphi(y,{\bf k})
$$
where
\begin{equation}
\int d^2 k\  K_{re}({\bf q},{\bf k})\varphi(y,{\bf k}) = 
{\bf q}^2\int d^2 k\ {1 \over {\bf k}^2 ({\bf q} - {\bf k})^2
} \varphi(y,{\bf k})
\label{breal}
\end{equation}
is the contribution of real gluon emission to the 
evolution in $\alpha_s \ln 1/x$ and
\begin{equation}
\int d^2kK_{virt}({\bf q},{\bf k}) \varphi(y,{\bf k}) = -
{{\bf q}^2\over 2}\int d^2 k 
{1 \over {\bf k}^2 ({\bf q} - {\bf k})^2}\varphi(y,{\bf q})
\label{bvirtual}
\end{equation}
is the correspoding virtual contribution which eventually leads (after
resummation) to the gluon reggeization \cite{Li}, \cite{DD}.

The solution of the BFKL equation exhibits a powerlike growth at small
$x$,
\begin{equation}
\varphi (x, {\bf k}) \sim ({1 \over x})^{4 {{\bar\alpha_s}} \ln 2}
\end{equation}
which leads to apparent violation of unitarity at very small $x$.  It
was argued a long time ago \cite{GLR} that eventually the
system should enter a new regime, where the rate of growth of the
gluon density slows down and eventually saturates, thus curing a
potential conflict with unitarity of the underlying scattering.  The
responsibility for the slow down and the saturation should lie with
the nonperturbative finite density effects which are left out entirely
from the BFKL evolution.  The restoration of unitarity in high energy
(density) limit of QCD remains an outstanding problem which remains
unsolved although several approaches are being explored in the
literature \cite{Li}, \cite{BKP}, \cite{Mueller}.

This paper is based on the approach first proposed by McLerran and
Venugopalan \cite{MV} in the context of ultrarelativistic heavy ion
collisions.  The idea in \cite{MV} is that there is a regime of
high density and weak coupling in which semiclassical methods should
apply.  It is therefore suggested that the leading small $x$ glue
structure of the nucleus is due to the classical gluon field which is
created by the random color charges of energetic on-shell partons.
The nonlinearities of the Yang - Mills equations exhibit themselves
already on this classical level and it is therefore possible that they
provide the necessary saturation mechanism at low $x$. It is to be
stressed that this approach is intrinsically nonperturbative since
both the color charge density and the classical glue fields (gluon
density) are not considered to be small when solving the classical
equations.

Originally only the valence quarks were considered as the sources of
the color field \cite{MV}. It turned out that the quantum corrections
to this approximation \cite{AJMV} are big at low $x$.  Physically this
is understandable since at asymptotically small $x$ the source for the
glue field should include not only the valence quarks but all ``on
shell'' partons with longitudinal momenta $p^+>xP^+$ (where $P^+$ is
the momentum of a hadron). 

Experimentallly it is known \cite{HERA} that the dominant contribution
to the proton structure function at small Bjorken $x$ is given by gluons.
Their relative contribution, compared to the valence quark contribution,
grows when $x$ becomes smaller.

At small $x$ therefore gluons give the main
contribution to the color charge density.  The typical corrections to
the leading order semiclassical results were of order of $\alpha_s \ln
1/x $ showing the necessity of resumming the leading $\ln 1/x$
contribution.

The Wilson Renormalization Group approach to this resummation was
developed in \cite{JKMW}. The idea is to generate an effective
Lagrangian for low $x$ fields by integrating out all quantum
fluctuations around the classical background with longitudinal momenta
$p^+>xP^+$. The low $x$ glue field then should be determined by
solving classical equations that follow from this effective
Lagrangian. The main effect of this integrating out procedure is the
change in the color charge density distribution in the effective
Lagrangian. It was also realized in \cite{JKMW} that when formulated
in this way, the effective Lagrangian approach should be applicable
not only to heavy ions but to ordinary hadrons as well.  The double
logarithmic asymtotics in $\ln q^2 \ln 1/x$ was considered within this
framework as the first step in \cite{JKMW} .  The validity of this RG
approach should not depend on whether the classical background glue
field is strong or not. It is most interesting therefore to apply it
to the strong field - high density situation.  However, before
plunging into the nonlinear realm of strong fields, it is natural to
ask whether the approach of \cite{JKMW} at a simple linear level
reproduces the BFKL evolution equation.  This is in fact a crucial
test for the whole semiclassical philosophy, the importance of which
was recognized already in \cite{MV}.  The answer to this question
turns out to be affirmative.

Derivation of BFKL evolution equation from the renormalization group
improved low $x$ effective action is the main subject of this paper.
We shall also comment on the relation of this approach to that of
Lipatov \cite{Li}.  The paper is organized as follows. In Section 2 we
describe in some detail the structure of the low $x$ effective action
and the renormalization group procedure. In Section 3 we calculate the
RG equation in the weak field limit and show that in this limit it
reduces to the BFKL equation for the DIS gluon distribution function.
In Section 4 we comment on the relation of our approach to that of
Lipatov and make some closing remarks.

\section {The effective action and the renormalization group}

The starting point of our approach is the following action given in
the Light Cone gauge $A^+ = 0$ 
\begin{eqnarray}
S&=&i\int d^2 x_t F[\rho ^a(x_t)]\\
&-& \int d^4 x {1\over 4}G^2 + {{i}\over{N_c}} \int d^2 x_t dx^-
\delta (x^-)
\rho^{a}(x_t) {\rm tr}T_a W_{-\infty,\infty} [A^-](x^-,x_t)\nonumber \\
\label{action}
\end{eqnarray}

Here $G^{\mu \nu} $ is the gluon field
strength tensor
\begin{equation}
G^{\mu \nu}_{a} = \partial^{\mu} A^{\nu}_{a} 
- \partial^{\nu} A^{\mu}_{a} + 
g f_{abc} A^{\mu}_{b} A^{\nu}_{c}\nonumber
\end{equation}
$T_a$ are the $SU(N)$ color matrices in the adjoint representation,
and $W$ is the path ordered exponential along the $x^+$ direction in
the adjoint representation of the $SU(N_c)$ group
\begin{equation}
W_{-\infty,\infty}[A^-](x^-,x_t) = P\exp \bigg[-ig \int dx^+
A^-_a(x^-,x_t)T_a \bigg] \end{equation}
Even though the functional $S[\rho, A]$
has an imaginary part, we will use the term action when referring to
it since the average of any gluonic operator $O(A)$ in the hadron is
calculated as 
\begin{equation} <O>={\int [D\rho^a][DA^\mu_a]O(A)\
\exp\{iS[\rho, A]\}\over \int [D\rho^a][DA^\mu_a] 
\exp\{iS[\rho, A]\}}
\label{average}
\end{equation}
The exponential of the imaginary part of the action
\begin{equation}
{\rm Im}\ S=\int d^2x_t F[\rho^a]
\end{equation}
can be thought of as a kind of ``free energy''.
The ``Bolzmann factor'' 
\begin{equation}
\exp\{-\int d^2x_t F[\rho^a]\}
\label{bolzmann}
\end{equation}
appearing in (\ref{average}) controls the statistical weight of a
particular configuration of the two dimensional color charge density
$\rho^a(x_t)$ inside the hadron.

Since the action (\ref{action}) is a little different than the one used in
\cite{MV}, \cite{JKMW} some clarifying remarks are in order here. 
In the McLerran - Venugopalan (MV)
action \cite{MV}, \cite{JKMW} the ``free energy'' was taken to be a 
gaussian of the form 
\begin{equation}
F[\rho(x_t)]= {{1}\over {2\mu^2}} \rho_a^{2}(x_t)
\end{equation}
The dimensional constant $\mu^2$ then has the meaning of the average
color charge density squared per unit area.  This form of the ``free
energy'' is valid as long as the charge density is large and the color
charges that build it up are randomly distributed in color space.

In this paper we prefer not to specify $F[\rho]$ explicitly, but
rather think of it as a general positive definite functional. This we
do for two reasons. First, for the purpose of derivation of the BFKL
equation we will not need the explicit form of $F$. Second, the low
$x$ renormalization group procedure that we employ, in the general
nonlinear case results in an equation of the form 
\begin{equation} 
{d\over dy}F[\rho]=\alpha \Delta[\rho]
\label{rge}
\end{equation}
so that the functional form of $F$ changes as one considers lower and
lower values of $x$. In fact our expectation is that (\ref{rge}) will
have a fixed point as $x\rightarrow 0$, thereby providing a mechanism
for unitarization.

Another element which is different in the action (\ref{action}) from
the MV action is the last term which involves the Wilson line along
the $x^+$ direction. The origin of this term is easy to understand.
It is a natural gauge invariant extension of the abelian coupling of
the glue field to the external color charge density $\rho^a$.  The
equation of motion that follows from this action is 
\begin{equation} 
D_\mu G^{\mu\nu}=J^+\delta^{\nu +}
\label{eom}
\end{equation}
with
\begin{equation}
J^+_a(x)={g\over{N_c}} 
\delta (x^-) 
\rho^{b}(x_t) 
{\rm tr}\Bigg[T_b W_{-\infty,x^+} [A^-]T_a W_{x^+,\infty} [A^-]\Bigg]
\label{current}
\end{equation}
Expanding this expression for the current to lowest order in the field
$A^-$ yields
\begin{equation}
J^+_a(x)=g\delta (x^-)\rho^{a}(x_t) 
\end{equation}
which is the form of the current used in \cite{MV}, \cite{AJMV}.  As
explained in \cite{MV}, this form is only valid in the gauge
$A^-(x^-=0)=0$. In more general gauges the current should satisfy the
covariant conservation condition
\begin{equation}
D^-J^+=0
\end{equation}
It is straightforward to check that our current (\ref{current}) is indeed 
covariantly conserved. This is a direct consequence of the fact that
the action (\ref{action}) in the light cone gauge $A^+=0$
is still gauge invariant under the residual gauge transformations with gauge
functions which 
do not depend on $x^-$ 
and vanish at $x^+\rightarrow\pm\infty$. Under an infinitesimal 
transformation with such
a gauge function $\lambda^a(x_t,x^+)$
we have from the Wilson line in the action 
\begin{equation}
0=\delta S_W=\int \lambda^a(D^-{\delta S\over\delta A^-_a})\equiv
\int \lambda^a(D^-J^+)^a
\end{equation}
Thus the gauge invariance is equivalent to covariant conservation of
 the current\footnote{The charge 
density $\rho(x_t)$ has the meaning of a transition matrix element.
Roughly speaking,
$$\rho^a(x_t)\sim <\Psi_i|\int dx^- j^a(x_t,x^-)|\Psi_f>$$
where $j^a(x_t,x^-)$ is the charge density operator of the energetic partons,
and $|\Psi_i>$ and $|\Psi_f>$ are their wave functions at 
$x^+\rightarrow-\infty$ 
and  $x^+\rightarrow \infty$ respectively.
It is therefore natural that $\rho(x_t)$ is gauge invariant under gauge 
transformations that vanish at infinity. Under $x^+$ independent gauge
transformations it transforms as an adjoint field.}.

Now, the MV approximation for calculating the path integral in 
eq.(\ref{average})
is to find the classical solution for the equations of motion eq.(\ref{eom})
at fixed $\rho$ , and then
to average over the charge density distribution with the ``Bolzmann weight''
eq.(\ref{bolzmann}). 
The classical solution for any fixed $\rho(x_t)$ has the structure
\begin{eqnarray}
A^-_{cl}&=&0\nonumber \\
A^i_{cl}\equiv b^i&=&\theta(x^-)\alpha^i(x_t)
\end{eqnarray}
The presence of the extra terms in our action as compared to ref.\cite{MV}
does not change the classical solution since the extra terms involve only
$A^-$ which vanishes on the classical solution anyway.
The quantum corrections to this classical approximation are large at small
longitudinal momenta. To resum these large corrections we follow 
the renormalization group
procedure described in \cite{JKMW}.

Let us introduce the following decomposition of the gauge field:
\begin{equation}
A^a_{\mu} (x) = b^a_{\mu} (x) + \delta A^a_{\mu} (x) + a^a_{\mu} (x)
\end{equation}
where $ b^a_{\mu} (x)$ is the solution of the classical equations of motion,
$ \delta A^a_{\mu} (x) $
is the fluctuation field containing longitudinal momentum modes 
$q^+$ such that $P^+_n < q^+ < P^+_{n-1}$ while $a$ is a soft field
with momenta $k^+<P^+_n$, with respect to which the effective action
is computed. The initial path integral is formulated with the longitudinal
momentum cutoff on the field $\delta A$, $q^+<P^+_{n-1}$.
The effective action for $a^\mu$ is calculated by integrating 
over
the fluctuations  $\delta A$. This integration is performed
within the assumption that the fluctuations 
are small as compared to the classical fields
 $b^a_{\mu}$. More quantitatively, at each step of this RG procedure
the scale $P^+_n$ is chosen such that $\ln {P^+_{n-1}\over P^+_n}>1$, but
 $\alpha_s\ln {P^+_{n-1}\over P^+_n}<<1$. 
Expanding the action
around the classical solution $b^a_{\mu} (x)$ and keeping terms of the 
first and second order in
$\delta A$ we get 
\begin{equation}
S= -{1\over 4}G(a)^2-{{1}\over{2}} \delta A_{\mu} [{\rm D}^{-1} (\rho)]^{\mu\nu} \delta A_{\nu} + 
ga^- \rho^\prime+O((a^-)^2)
+ iF[\rho]
\label{effectiveaction}
\end{equation}
where
\begin{equation}
\rho^\prime =\rho+ \delta \rho_1 + \delta \rho_2
\label{prime}
\end{equation}
with
\begin{eqnarray}
\delta \rho^a_1(x_t,x^+) &=& -2 f^{abc} \alpha^{b}_{i}\delta A^{c}_{i}(x^-=0)
\label{rho1}
\\ 
&-& {{g}\over{2}} f^{abc} \rho^{b}
(x_t) \int dy^+ \Bigg[\theta (y^+ - x^+) - \theta (x^+ - y^+) \Bigg] 
\delta A^{-c}(y^+,x_t, x^-=0)\nonumber
\end{eqnarray}
and
\begin{eqnarray}
\delta \rho^a_2(x) &=& -f^{abc} [\partial^+ \delta A^{b}_{i}(x) ]\delta A^{c}_{i}(x) \nonumber\\
&-& 
{{g^2}\over{N_c}} \rho^{b}(x_t) \int dy^+ \delta A^{-c}(y^+,x_t,x^-=0) \int dz^+ \delta A^{-d}(z^+,x_t,x^-=0) \nonumber \\
&\times &\Bigg[\theta (z^+ -y^+) \theta (y^+ -x^+)  {\rm tr} T^a T^c T^d T^b  \nonumber \\
&+&\theta (x^+ -z^+) \theta (z^+ -y^+)  {\rm tr} T^a T^b T^c T^d  \nonumber \\
&+&\theta (z^+ -x^+) \theta (x^+ -y^+)  {\rm tr} T^a T^d T^b T^c \Bigg]
\label{rho2}
\end{eqnarray}
The first term in both $\delta\rho_1^a$ and 
$\delta \rho^a_2$ is coming from expansion of $G^2$ in 
the action
while the rest of the terms proportional to $\rho$
are from the expansion of the 
Wilson line term. 
The three terms
correspond to different time ordering of the fields. 
Since the longitudinal momentum of $a^-$ is much lower than of $\delta A$, we 
have only kept the eikonal coupling (the coupling to $a^-$ only), which gives
the leading contribution in this kinematics.
The inverse propagator $[{\rm D}^{-1}]^{\mu\nu}$ is given by
\begin{eqnarray}
\left[{\rm D}^{-1}\right]^{ij}_{ab}(x,y)
&=&\left[D^2(b)\delta^{ij}+D^i(b)D^j(b)\right]_{ab}\delta^{(4)}(x,y)
\\
\left[{\rm D}^{-1}\right]^{i+}_{ab}(x,y)
&=&-[\partial^+_xD^i_{ab}(b)\delta^{(4)}(x,y)+2f_{abc}\alpha_c^i(x_t)
\delta(x^-)\delta(y^-)\delta^{(2)}(x_t,y_t)\delta(x^+,y^+) ]\nonumber\\
\left[{\rm D}^{-1}\right]^{++}_{ab}(x,y)&=&(\partial^+)^2\delta_{ab}\delta^{(4)}(x,y)+
f_{abc}\rho^c(x_t)\delta(x^-)\delta(y^-)\theta(x^+-y^+)\delta^{(2)}(x_t,y_t)
\nonumber
\end{eqnarray}

Two things should be noted here. First, formally $\delta\rho$ is a function
of $x^+$ and $x^-$ as well as $x_t$. However, it is a function
of $\delta A$'s which only have longitudinal momenta much larger than the 
momenta in the soft field $a$. 
The (light cone) time variation scale of $\delta\rho$ is 
therefore ${1\over q^-}\sim{q^+\over q_t^2}$ and is much larger than the 
typical time variation scale of the on shell modes of the field $a$. From this
point of view $\delta \rho$ is therefore for all practical purposes 
(light cone) time independent. 
As for $x^-$ dependence, the only term which
does not have an explicit $\delta(x^-)$ is the first term in eq.(\ref{rho2}).
However remember that we are only interested in its low longitudinal
momentum components since it couples directly to $a^-$ in 
the effective action. So, in momentum space we are interested in
$f^{abc} \int dq^+[q^+ \delta A^{b}_{i}(q^+) ]
\delta A^{c}_{i}(-q^++k^+)$. 
Since the leading logarithmic contributions comes from the region
$q^+>>k^+$, to this accuracy this expression does not depend on $k^+$ and is
therefore also $\delta(x^-)$ in coordinate space.
Therefore $\delta\rho$  behaves as an honest-to-God addition to the
static, local in $x^-$ charge density $\rho$. 

Second, note that we have not written out explicitly higher order in 
$a^-$ terms in the effective action. There are of course such terms, which 
come from expanding the Wilson line part of the action. Disregarding
these terms gives the effective action with the coupling of 
the field $a^-$
to the charge density of the form $a^-\rho$.
However, imposing gauge invariance on the final result together with the
requirement that the linear in $a^-$ term of the gauge invariant action should
coincide with the result of our calculation, the full gauge invariant
form of the effective action will be recovered. In the following therefore
we will concentrate on the linear term $a^-\rho$ only.

The procedure now is the following:

1. Integrate over $\delta A^\mu$ at fixed $\rho$ and fixed $\delta\rho$.

2. Integrate over $\rho$ at fixed $\rho^\prime=\rho+\delta\rho$.

This generates the new effective action which formally can be written as
\begin{equation}
\exp\{iS[\rho^\prime, a^\mu]\}=\exp\{-F^\prime[\rho^\prime]-{i\over 4}G^2(a)
+iga\rho^\prime\}
\end{equation}
with
\begin{equation}
\exp\{-F^\prime[\rho^\prime]\}=
\int [D \rho ][D \delta A]\  \delta (\rho^\prime-\rho-\delta \rho[\delta A])
 \exp\{-F[\rho]
-{i\over 2} \delta A D^{-1} [\rho] \delta A\}
\label{newf}
\end{equation}
Of course, to leading order in $\ln 1/x$ only terms linear in $\alpha_s\ln 1/x$
should be kept in $F^\prime$. Defining
\begin{equation}
\alpha_s\ln {1\over x}\Delta[\rho]\equiv F^\prime[\rho]-F[\rho]
\end{equation}
gives the RG equation
\begin{equation}
{d\over dy}F[\rho]=\alpha \Delta[\rho]
\label{evol}
\end{equation}

In this way our renormalization group procedure leads to a set of
evolution equations
for all coefficient functions in the functional $F[\rho]$. 
Technically the most complicated part of the calculation in eq.(\ref{newf})
turns out to be the inversion of the operator $D^{-1}$ which arises after
integration over $\delta A$.
The rest of the 
integrations in eq.(\ref{newf}) can be performed explicitly and it is possible
to express $\Delta$ explicitly in terms of $D[\rho]$.
This work is now in progress. 
In this paper however, we will only consider the weak field limit. In this
limit $D$ can be expanded in powers of $\rho$. We will show in the next section
that the evolution equations eq.(\ref{evol}) yield in this limit the celebrated
BFKL equation.

\section{The weak field limit and the BFKL equation}

In our semiclassical approach the glue distribution function is given by
(see ref.\cite{MV})
\begin{equation}
g(x,Q^2)={1\over x}\int_0^{Q^2} d^2k_td^2(x_t-y_t)e^{-ik_t(x_t-y_t)}
<<\alpha^i_a(x_t)\alpha^i_a(y_t)>>_\rho
\end{equation}
where by $<< \ \  >>_\rho$ we denote the average over the ensemble of $\rho(x_t)$
with the statistical weight eq.(\ref{bolzmann}).
In the weak field regime the classical field $\alpha^i$ is related to the
charge density by (\cite{MV})
\begin{equation}
\alpha^i_a=-{\partial^i\over \partial_t^2}\rho_a
\end{equation}
It therefore follows that
\begin{equation}
xg(x,Q^2)=\int_0^{Q^2} {d^2k_t\over k_t^2}\varphi(k_t)
\end{equation}
with
\begin{equation}
\varphi(k_t)=\int d^2(x_t-y_t)e^{-ik_t(x_t-y_t)}
<<\rho(x_t)\rho(y_t)>>_\rho
\label{iden}
\end{equation}
The unintegrated gluon density $\varphi$ is therefore nothing but the
Fourier transform of the correlation function of the color charge density.
To see how the BFKL kernel arises we should therefore compute the charge 
density correlation function 
$<<\rho^\prime\rho^\prime>>_\rho$ after one step in our RG procedure.

One could do it of course by calculating $F^\prime$ first, and 
then averaging over $\rho$. However in the weak field case it is simpler
to directly express $<<\rho^\prime\rho^\prime>>$ 
in terms of  $<<\rho\rho>>$ using the explicit relations
eqs.(\ref{prime},\ref{rho1},\ref{rho2})
\begin{eqnarray}
<< \rho^\prime \rho^\prime>>_\rho& =& 
<< [<(\rho + \delta\rho )( \rho + \delta \rho )>_ {\delta A}]>>_\rho\\
& =& 
<<\rho \rho>>_\rho + 2<< \rho <\delta \rho>_{\delta A} >>_\rho
+<<[<\delta \rho \delta \rho >_{\delta A}]>>_\rho \nonumber
\label{rhoprimecorelator}
\end{eqnarray}
Here $< \ \ >_{\delta A}$ denotes averaging over the hard fluctuation field
$\delta A$ at fixed value of $\rho$.
The relation between the correlators becomes explicit already after
averaging over $\delta A$ so that the averaging over $\rho$ does not need
to be performed explicitly.

Noting that $\delta \rho_1$ is linear
in $\delta A$ while $\delta \rho_2$ is quadratic in $\delta A$  
(eqs.\ref{rho1},\ref{rho2}) we have
\begin{equation}
<\delta \rho >_{\delta A} = < \delta \rho_2>_{\delta A}
\label{virtual}
\end{equation}
and
\begin{equation}
<\delta \rho \delta \rho >_{\delta A} = <\delta \rho_1 \delta \rho_1>_{\delta A}
\label{real}
\end{equation}
The second equation is valid to leading order in $\alpha_s\ln {1\over x}$.
As we shall see now, eqs.(\ref{virtual}) and (\ref{real}) contribute the
virtual and the real part of the BFKL kernel respectively.

\subsection{The Virtual Part}
Let us start by considering eq.(\ref{virtual}).
It has the diagrammatic representation of Fig.1.

\begin{center}
\begin{picture}(375,50)
\linethickness{3pt}
\multiput (0,0)(100,0){4}{\line(80,0){80}}
\thicklines
\multiput(40,-24)(0,-5){6}{\line(0,1){1}}
\multiput(140,0)(0,-5){6}{\line(0,1){1}}
\multiput(210,0)(0,-5){6}{\line(0,1){1}}
\multiput(372,0)(0,-5){6}{\line(0,1){1}}
\thinlines
\GlueArc(40,0)(20,180,360){2}{10}
\GlueArc(140,0)(20,0,180){2}{10}
\GlueArc(240,0)(20,0,180){2}{10}
\GlueArc(340,0)(20,0,180){2}{10}
\put(35,-65){1.a}
\put(135,-65){1.b}
\put(235,-65){1.c}
\put(335,-65){1.d}
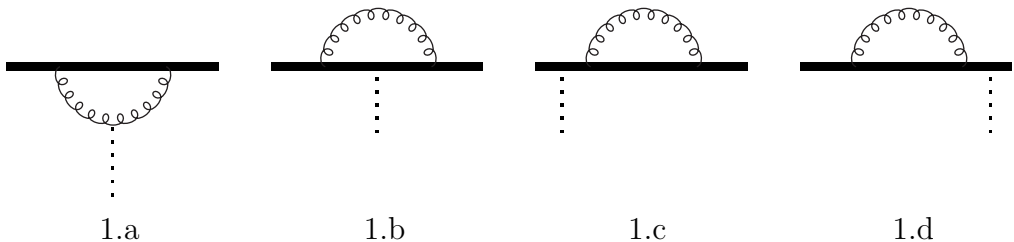
\put(0,-150){\parbox[b]{5.5in}{Figure 1: \it{The virtual diagrams. 
The solid lines represent the effective vertex arising 
from expansion of the Wilson line in the action. The wavy line
represents the free propagator of the hard fluctuation field $\delta
A$. The dashed lines indicate the coupling of the soft fluctuation
field $a^{-}$.}}} 
\end{picture}
\end{center}
\vspace{2.3in}

The diagram on Fig. 1a corresponds to the calculation of
\begin{equation}
<\delta\rho^a_{21}(x)>_{\delta A}
\equiv -g f^{abc} 
< (\partial^+ \delta A^{b}_{i} \delta A^{c}_{i} >_{\delta A}
\end{equation}
where the propagator of the fluctuation field is expanded to first
order in the color charge density $\rho$.  Explicitly\footnote{There
  is in principle another contribution to this quantity: the standard
  gluon self energy diagram. It however does not give a contribution
  logarithmic in $x$, and for this reason is not considered here.}
\begin{eqnarray}
< (\partial^+ \delta A^{b}_{i}(x) \delta A^{c}_{i}(x) >_{\delta A} 
&=&
\int [D\delta A_{i}] (\partial^+ \delta A^{b}_{i}(x)) \delta
A^{c}_{i}(x) 
\exp [-{{i}\over {4}} G^2 ] 
\\ &\times& \nonumber 
\Bigg[{{g^2}\over {N_c}} \int d^4 z_t  \rho^{a} (z_t)  
{\rm tr} T^a \int^{z^+} dy^+ \delta A^{-}
(y^+) \delta A^{-}(z^+) \Bigg]
\end{eqnarray}
In terms of free gluon propagators this is
\begin{eqnarray}
< (\partial^+ \delta A^{b}_{i}(x) \delta A^{c}_{i}(x) >_{\delta A}
&=& {{ig^2}\over{2}} f^{abc} \int d^4 z
 \rho^{a} (z_t) \nonumber \\
&\times& \Bigg[ - \partial^{+}_{x} G^{i-} (x^+ - z^+) 
\int^{z^+} dy^+ G^{i-}(x^+ -y^+) \nonumber \\
&+& \qquad  \int^{z^+} dy^+ \partial^{+}_{x} 
G^{i-} (x^+ -y^+) G^{i-} (x^+ -z^+) \Bigg]
\end{eqnarray}
where only the $+$ components of the arguments of the propagators are 
shown explicitly since 
the transverse coordinates are all equal, and $y^-=z^-=0$.
Here
\begin{eqnarray}
G^{--}(p)&=&{2ip^-\over p^+ p^2}\\
G^{-i}(p)&=&{ip_i\over p^+p^2}\nonumber
\end{eqnarray}
Performing the  $z^+$, $z^-$, $y^+$ and $y^-$ integrations we obtain
\begin{eqnarray}
< (\partial^+ \delta A^{b}_{i}(x) \delta A^{c}_{i}(x) >_{\delta A} 
&=& -i g^2 f^{abc} \int d^2 z_t \rho^{a}
(z_t) \int {{d^2 q_t}\over{(2\pi)^2}} {{d^4 p}\over{(2\pi)^4}} 
e^{i(p_t + q_t )( x_t - z_t)}
\nonumber \\
& &\qquad \times 
\Bigg[{{p_t \cdot q_t}\over{p^+ (p^- - i\epsilon )
(p^2 + i\epsilon )(2p^-p^+ - q_{t}^{2} 
+ i\epsilon )}}\Bigg]
\nonumber
\end{eqnarray}
The $i\epsilon$ prescription for the $p^-$ pole follows from adding a
convergence factor to the exponentials when performing integrals of
the form
\begin{equation}
\int^{z^+}_{-\infty} dy^+ e^{ip^- y^+}
\nonumber
\end{equation}
so that there is no contribution from infinity. 
Also the $p^+$ integral is restricted to run
between the lower and upper limits of the momentum region where 
the hard fluctuation fields
are being integrated out. 

Closing the integration contour in the complex $p^-$ plane, the
$p^-$ integration is easily performed. The $p^+$ integral then
factors out as
$\int^{P^+_{n-1}}_{P^+_n} {{dp^+}\over{p^+}} $  which gives 
$\ln {P^+_{n-1}\over P^+_n}\equiv \ln {x_{n-1}\over x_{n}}$.

Finally
\begin{equation}
< (\partial^+ \delta A^{b}_{i}(x) \delta A^{c}_{i}(x) >_{\delta A} = 
{{g^2}\over{2\pi}} f^{abc} 
 \ln {x_{n-1}\over x_{n}} \int d^2 z_t \rho^{a}(z_t)  \int {{d^2 q_t}\over{(2\pi)^2}} 
{{d^2 p_t}\over{(2\pi)^2}} {{p_t \cdot q_t}\over{p_t^2 q_t^2}}  
e^{i(p_t + q_t )( x_t - z_t)}
\end{equation}
Fourier
transforming the above equation and shifting $p_t \rightarrow p_t - q_t $ 
gives
\begin{equation}
< \delta \rho^{a}_{21}(k_t) >_{\delta A} = - {{g^2 N_c}\over{(2\pi)^3}} 
 \ln {x_{n-1}\over x_{n}} \rho^{a}(k_t)
\Bigg[ {{k_t^2}\over{2}} \int {{d^2 p_t}\over {p_t^2 (p_t - k_t )^2}} - 
\int {{d^2 p_t}
\over {p_t^2}}\Bigg]
\label{virtual1}
\end{equation}

The contribution of Figs. 1(b,c,d) to the color charge density is
\begin{eqnarray}
<\delta\rho^{a}_{22}(x_t) >_{\delta A} &=& -g^2 N_c \rho^{a}(x_t)\int dy^+ dz^+ 
G^{--} (y^+ - z^+)
\Bigg[\theta ( z^+ - y^+)\theta (y^+ - x^+)\nonumber \\ 
&+& \theta ( x^+ - z^+)\theta (z^+ - y^+) + 
{{1}\over{2}}\theta ( z^+ - x^+)\theta (x^+ - y^+)  \Bigg]
\end{eqnarray}
This is the average of the rest of the terms in eq.(\ref{rho2})
with all gloun propagators taken to be free propagators.
The different theta functions represent 
the different order of emision of the soft 
field $a^{-}(x^+)$. The first two terms correspond to 
diagrams where the soft field
is emitted either before or after the hard fluctuation has been emitted and
reabsorbed, while the last term 
corresponds to emission during the time the hard fluctuation was 
``in flight''. The factor $1/2$ in the third term is the result of 
color algebra.
Using the identity
\begin{equation}
\theta ( z^+ - y^+)\theta (y^+ - x^+) + \theta ( x^+ - z^+)
\theta (z^+ - y^+) + 
\theta ( z^+ - x^+)\theta (x^+ - y^+) = \theta (z^+ - y^+)
\nonumber
\end{equation}
this expression can be written as
\begin{eqnarray}
<\delta\rho^{a}_{22}(x_t) >_{\delta A} & = & -g^2 N_c \rho^{a}(x_t) 
\int dy^+ dz^+ G^{--} (y^+ - z^+)
\nonumber \\ && \qquad \times
\bigg[ \theta (z^+ - y^+) -{
\theta ( z^+ - x^+)\theta (x^+ - y^+)\over{2}}  \bigg]
\end{eqnarray}
The first term which is totally independent of $x^+$, vanishes
since it is proportional to $ G^{--} (p^- = 0)$ and 
$G^{--} (p^{-} ) \sim {{p^-}\over{p^+ p^2}}$
We therefore obtain
\begin{equation}
<\delta\rho^{a}_{22}(x_t) >_{\delta A} = {{g^2 N_c}\over{2}} \rho^{a}(x_t) \int dy^+ dz^+ 
G^{--} (y^+ - z^+) \theta ( z^+ - x^+)\theta (x^+ - y^+)
\end{equation}
After Fourier transforming to momentum space and using the explicit form of
the free propagator for $G^{--}$ this yields
\begin{equation}
<\delta\rho^{a}_{22}(k_t) >_{\delta A} = -{{g^2 N_c}\over{(2\pi)^3}} \ln {x_{n-1}\over 
x_{n}}
\rho^{a}(k_t) \int {{d^2 p_t}\over{p_t^2}}
\label{virtual2}
\end{equation}
Combining equations~(\ref{virtual1}) and~(\ref{virtual2}) gives
\begin{equation}
<\delta\rho^{a}_{2}(k_t) >_{\delta A} = -{{g^2 N_c}\over{2(2\pi)^3}}
\ln {x_{n-1}\over x_{n}}
\rho^{a}(k_t) \int d^2 p_t {{k_t^2}\over{p_t^2 (p_t - k_t)^2}}
\label{totvir}
\end{equation}
This is the contribution of all virtual diagrams to the change in the color
charge density.   

\subsection{The Real Part}
Now we have to compute the connected correlator of $\delta\rho_1$
\begin{equation}
\delta \rho^{a} = -2 f^{abc} \alpha^{b}_{i} \delta A^{c}_{i} - {{g}\over{2}}
f^{abc} \rho^{b} (x_t) \int dy^+ \bigg[ \theta (y^+ - x^+) - 
\theta (x^+ - y^+)\bigg] \delta A^{-c} (y^+)
\label{rho1a}
\end{equation}
The two theta functions in the second term 
again correspond to emmision of the soft field before and after the 
emmision of the hard fluctuation respectively. 

The procedure to compute $<\delta \rho^{a}_1 (x_t) \delta \rho^{a}_1
(y_t) >_{\delta A} $ is as before; we contract the hard fluctuation
fields $\delta A$ and use free propagators for $<\delta A \delta A
>_{\delta A}$. Squaring the first term in eq. (\ref{rho1a}) gives
\begin{equation}
<\delta \rho^{a}_{11} (x_t) \delta \rho^{a}_{11} (y_t) >_{\delta A} = 
{{2N_c}\over{(2\pi )^3}}
(2\pi )^2 \delta^2 (x_t - y_t)  \ln {x_{n-1}\over x_{n}} \alpha^{a}_{i} (x_t) 
\alpha^{a}_{i}(y_t)
\label{oneone}
\end{equation}
This contribution diagrammatically is depicted on Fig. 2.a.

\begin{center}
\begin{picture}(375,50)
\linethickness{4pt}
\thicklines
\put(100,0){\line(2,-3){10}}
\put(110,0){\line(-2,-3){10}}
\linethickness{1pt}
\put(105,-8){\line(0,-40){40}}
\linethickness{4pt}
\put(200,0){\line(2,-3){10}}
\put(210,0){\line(-2,-3){10}}
\linethickness{1pt}
\put(205,-8){\line(0,-40){40}}
\Gluon(105,-48)(205,-48){3}{13}
\multiput(105,-48)(0,-5){8}{\line(0,1){1}}
\multiput(205,-48)(0,-5){8}{\line(0,1){1}}
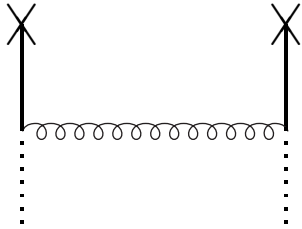
\put(0,-140){\parbox[b]{5.5in}{Figure 2.a: \it{The diagram for the 
contribution of eq.~(\ref{oneone}). The crosses denote insertions of the 
background field $\alpha_i$. The rest of the notations are as in Fig.1.}}}
\end{picture}
\end{center}
\vspace{1.8in}

Squaring the second term (Fig. 2.b) gives
\begin{equation}
<\delta \rho^{a}_{12} (x_t) \delta \rho^{a}_{12} (y_t) >_{\delta A} = 
{{2g^2 N_c}\over{(2\pi )^3}}
 \ln {x_{n-1}\over x_{n}} 
\rho^{a} (x_t)\rho^{a} (y_t) \int {{d^2 p_t}\over{p_t^2}} 
e^{ip_t (x_t - y_t)}
\label{twotwo}
\end{equation}
\begin{center}
\begin{picture}(375,50)
\linethickness{3pt}
\multiput (0,0)(45,0){2}{\line(30,0){30}}
\multiput (110,0)(45,0){2}{\line(30,0){30}}
\multiput (220,0)(45,0){2}{\line(30,0){30}}
\multiput (330,0)(45,0){2}{\line(30,0){30}}
\thinlines
\GlueArc(37.5,0)(20,0,180){2}{10}
\GlueArc(147.5,0)(20,0,180){2}{10}
\GlueArc(249,0)(20,0,180){2}{10}
\GlueArc(372,0)(20,0,180){2}{10}
\thicklines
\multiput(8,0)(0,-5){8}{\line(0,1){1}}
\multiput(67,0)(0,-5){8}{\line(0,1){1}}
\multiput(135,0)(0,-5){8}{\line(0,1){1}}
\multiput(160,0)(0,-5){8}{\line(0,1){1}}
\multiput(240,0)(0,-5){8}{\line(0,1){1}}
\multiput(285,0)(0,-5){8}{\line(0,1){1}}
\multiput(340,0)(0,-5){8}{\line(0,1){1}}
\multiput(380,0)(0,-5){8}{\line(0,1){1}}
\thinlines
\put(0,-100){\parbox[b]{5.5in}{Figure 2.b: \it{Contribution of
eq.~(\ref{twotwo}).}}}
\end{picture}
\end{center}
\vspace{1.5in}

while the contribution of the cross term (Fig. 2.c) is
\begin{equation}
2<\delta \rho^{a}_{12} (x_t) \delta \rho^{a}_{21} (y_t) >_{\delta A} = 
{{4ig N_c}\over{(2\pi )^3}}
 \ln {x_{n-1}\over x_{n}} \alpha_{i}^{a} (x_t)\rho^{a} (y_t) \int {{d^2 p_t}\over
{p_t^2}} p_{i}  e^{ip_t (x_t - y_t)}
\label{onetwo}
\end{equation}

\begin{center}
\begin{picture}(375,50)
\linethickness{4pt}
\thicklines
\put(0,0){\line(2,-3){10}}
\put(10,0){\line(-2,-3){10}}
\linethickness{1pt}
\put(5,-8){\line(0,-40){40}}
\multiput(5,-48)(0,-5){8}{\line(0,1){1}}
\linethickness{4pt}
\thicklines
\put(230,0){\line(2,-3){10}}
\put(240,0){\line(-2,-3){10}}
\linethickness{1pt}
\put(235,-8){\line(0,-40){40}}
\multiput(235,-48)(0,-5){8}{\line(0,1){1}}
\linethickness{3pt}
\put(70,-8){\line(50,0){50}}
\put(300,-8){\line(50,0){50}}
\linethickness{1pt}
\multiput(110,-8)(0,-5){16}{\line(0,1){1}}
\multiput(310,-8)(0,-5){16}{\line(0,1){1}}
\thinlines
\Gluon(5,-48)(100,-9){3}{13}
\Gluon(235,-48)(345,-9){3}{15}
\put(0,-130){\parbox[b]{5.5in}{Figure 2.c: \it{Contribution of
eq.~(\ref{onetwo}).}}}
\end{picture}
\end{center}
\vspace{1.8in}

Fourier transforming to momentum space, using
\begin{equation}
\alpha^{a}_{i} (k_t) = -ig {{k_i}\over {k_t^2}} \rho^{a} (k_t)
\end{equation}
and combining the three contributions we obtain
\begin{eqnarray}
<\delta \rho^{a}_1 (k_t) \delta \rho^{a}_1 (-k_t) >_{\delta A} &=& 
{{2g^2 N_c}\over{(2\pi )^3}}
 \ln {x_{n-1}\over x_{n}} \int d^2 p_t \rho^{a} (p_t)\rho^{a} (-p_t)
\nonumber \\
& &\Bigg[ {{1}\over{p_t^2}} + {{1}\over {(p_t - k_t)^2}} - 2 {{p_t \cdot 
(p_t - k_t )}\over {p_t^2 (p_t - k_t)^2}} \Bigg]
\end{eqnarray}
The use of the identity
\begin{equation}
-2 {{p_t \cdot (p_t - k_t)}\over{p_t^2 (p_t - k_t)^2 }} = - {{1}\over{p_t^2}} 
- {{1}\over{(p_t - k_t)^2}} + {{k_t^2}\over{p_t^2 (p_t - k_t)^2}}
\end{equation}
gives
\begin{eqnarray}
<\delta \rho^{a}_1 (k_t) \delta \rho^{a}_1 (-k_t) >_{\delta A} &=& 
{{2g^2 N_c}\over{(2\pi )^3}}
 \ln {x_{n-1}\over x_{n}} \int d^2 p_t \rho^{a} (p_t)\rho^{a} (-p_t) {{k_t^2}
\over{p_t^2 (p_t - k_t)^2}}
\label{totreal}
\end{eqnarray}
This is the contribution of all real diagrams to the change in charge density.

Inserting  the expressions eqs.(\ref{totreal}) and (\ref{totvir}) into 
equation~(\ref{rhoprimecorelator}) we obtain
\begin{eqnarray}
\lefteqn{<<\rho^\prime(k_t) \rho^\prime(-k_t) >>_\rho 
  - <<\rho(k_t) \rho(-k_t)>>_\rho}\nonumber \\
&=& - 
{{g^2 N_c}\over{(2\pi)^3}}  \ln {x_{n-1}\over x_{n}}\int d^2 p_t 
{{k_t^2}\over{p_t^2 (p_t - k_t)^2}}
\nonumber \\&& \qquad \times
\Bigg[<<\rho^{a} (k_t) \rho^{a} (-k_t)>>_\rho - 2 <<\rho^{a} (p_t) 
\rho^{a} (-p_t)>>_\rho \Bigg]
\label{finitebfkl}
\end{eqnarray}
Identifying $<<\rho(k_t) \rho(-k_t)>>_\rho=\varphi(y, k_t)$, and
$<<\rho^\prime(k_t) \rho^\prime(-k_t) >>_\rho=\varphi(y+dy, k_t)$ as
in eq. (\ref{iden}), the equation (\ref{finitebfkl}) can be rewritten
in differential form to give precisely the BFKL equation
eq.(\ref{BFKL}).

\section{Discussion}
We have shown in this paper that the RG equation of our low $x$ Wilson
renormalization group procedure reduces to the BFKL equation in the
weak field limit.  This in itself is very satisfying and encourages
us to continue the study of our low $x$ effective action.

It also demonstrates the relation of the semiclassical aproach of
\cite{MV}, \cite{JKMW} in its present form to other more conventional
approaches to low $x$ physics.  In this connection it is specially
illuminating to compare it with the effective action approach of
Lipatov \cite{li2}. In fact, an attentive reader must have noticed a
lot of similarities between our effective action eq.  (\ref{action})
and the one used by Lipatov for description of the high energy
scattering in the multi-Regge limit of QCD\cite{Li},\cite{li2}
(cf. Eqs.~210, 249 in \cite{Li}).

\be
S_{Lipatov}&=&\partial_{\mu} R^-_a \partial^{\mu} R^+_a
              -{1\over 4}G^2\\
&+& {{i}\over{N_c}}
\partial^2 R^+_{a}(x) {\rm tr}T_a W_{x^+=-\infty,x^+=\infty} [A^-]\nonumber \\
&+& {{i}\over{N_c}}
\partial^2 R^-_{a}(x) {\rm tr}T_a W_{x^-=-\infty,x^-=\infty} [A^+]\nonumber \\
\label{liaction}
\ee
where
\begin{equation}
\partial^+R^-=\partial^-R^+=0
\end{equation}
This action describes gluons $A^\mu$ interacting with reggeons $R^+$
and $R^-$ corresponding to the quasistatic gluon fields
dominating the $t$ - channel exchange in the 
high energy scattering in multiregge
kinematics.
Our static color charge density obviously plays the role of the reggeon fields.
More explicitly, by comparing the interaction terms in (\ref{action}) and
(\ref{liaction}) one should identify $\rho(x_t) \delta(x^-)$ with
$\partial^2 R^+$.
The Wilson line term, which describes the
interaction of the reggeon with gluons is otherwise the same in
eqs.(\ref{action}) and (\ref{liaction}).  The calculational strategy
is actually similar up to a point.  In both approaches one first solves
for a gluon field in a given reggeon (classical charge)
background and then expands the
action around this classical configuration.

There are however some conceptual as well as technical differences
between the two approaches. As far as we can tell the reason for this is 
twofold. First our action is constructed to describe the structure of
one source of color field (nucleus or hadron DIS structure function) 
and not the collision of two hadrons.
In our case only one component of the reggeon field (charge
density) does not vanish, and $R^-=0$.  Secondly although one is tempted at
first to identify the $\partial_{\mu} R^+_a \partial^{\mu} R^-_a$
(reggeon
propagator) term in eq.(\ref{liaction}) with our $F[\rho]$, the two
terms in fact represent very different physics.  As discussed in
Section 2, $F[\rho]$ is the statistical weight with which a given
charge density (reggeon) configuration is present in the hadronic wave
function. Consequently, it appears as an {\it imaginary} part of the
action. The reggeon propagator term in $S_{Lipatov}$ on the other hand
is real and is included to reproduce the Born term in the scattering
amplitude of one hadron in the Coulomb field of the other\footnote{ In
  fact it seems to us that besides the reggeon kinetic term 
the weight factor for different reggeon
  configurations should be present also in the two hadron case.}.
Related to this is the fact, that technically the BFKL kernel appears
quite differently in the two calculations. In our case the action
calculated on the solution of classical equations of motion vanishes.
The contribution to the kernel therefore entirely comes from the small
fluctuation integral.  In the case of $S_{Lipatov}$, \cite{li2} it is
only the virtual part of BFKL kernel that comes from the integral over
the small fluctuation.  The real part of the kernel according to
\cite{li2} appears already in the action calculated on the classical
solution.

We would also like to note, that due to the fact that in our approach
$J^-$ vanishes, the classical solution is simpler. We are hoping
therefore that the small fluctuation propagator in this classical
background can be calculated explicitly along the lines of
\cite{AJMV}.

The low $x$ renormalization group procedure described in the present
paper and in \cite{JKMW} is a new element and it would be also
interesting to incorporate it in the analysis of $S_{Lipatov}$. 
Here we want to make
one remark about the nature of this renormalization group procedure.
It looks different from the standard RG used in the analysis of DIS at
moderate $x$, in that it is the change in the longitudinal rather than
the transverse momentum scale that defines the block spin procedure.
Conceptually however, the two are very similar.  It is more physical
to think about the modes which are being integrated out not in terms
of spatial momentum, but rather in terms of the frequency $p^-$. The
leading contribution to the scattering cross section comes from the
interaction of the external probe with the quark and gluon modes which
are practically static during the interaction time. The influence of
the modes with frequencies higher than the inverse interaction time of
the scattering process averages to zero. When the interaction time
becomes smaller, more and more field modes behave like static fields
from the viewpoint of the external probe. The intensity of the
``static'' field therefore effectively grows as the time of the
interaction decreases.  

Both lowering $x$ at fixed $Q^2$ and raising
$Q^2$ at fixed $x$ in DIS result in shorter interaction time between
the hadron and the external probe.  This is precisely the physical
mechanism which underlies the growth of the partonic distributions
both at smal $x$ and at large $Q^2$.  Roughly one can think of the on
shell frequency $p^-={p^2_t\over p^+}$ as defining the relevant time
scale for quantum fluctauations with given spatial momentum.  When the
longitudinal momenta of all the relevant modes are of the same order
the frequency is determined entirely by the transverse momentum. The
evolution in frequency then becomes the evolution in $Q^2$. This is
the situation in the DIS at moderate $x$. If all the transverse
momenta are of the same order, the frequency is governed by the
longitudinal momentum. This is the case of the low $x$ renormalization
group discussed in this paper. 

The interchangeability of $p^-$ and $1/p^+$
breaks down however, if the interaction spreads over wide range of
transverse momenta. This seems to be the case in the BFKL evolution,
the asymptotic solution for which has the character of a random walk
in the transverse momentum space \cite{askew}.  It would be interesting
to formulate the unifying RG procedure, which uses the frequency
directly as the RG evolution parameter. Potentially this could cure
the problem with the low $k_t$ modes in the BFKL evolution, since in
this kind of procedure those, being low frequency modes should apear
in the initial condition rather than in the evolution itself.

To summarize, in this paper we have considered the renormalization
group approach to the low $x$ effective action in the weak field
regime.  Clearly the more interesting and more complicated part of the
problem is to penetrate the nonlinear regime, where the fluctuation
field propagator has to be computed to all orders in $\rho$. This work
is now in progress.

{\bf Acknowledgements}
We are very grateful to Larry McLerran for numerous discussions on a variety
of topics related to the subject of this paper and for his constant 
encouragement and interest in this work. We have also benefitted from 
discussions with M. Gyulassy,  Yu. Kovchegov, 
A. Makhlin, B. Mueller, J.-W. Qiu, R. Rodriguez, E. Surdutovich, R. Venugopalan and 
G. Zinovjev. We thank INT, University of Washington 
for hospitality and financial 
support during the program ``Ultrarelativistic Nuclei: 
from structure functions to the quark-gluon plasma'' where part of this work
was done.
The work of A.L. was partially supported by Russian Fund for Basic Research,
Grant 96-02-16210.

\end{document}